\documentclass[a4paper]{article}
\usepackage[utf8]{inputenc}

\usepackage[english]{babel}

\usepackage{times}

\usepackage{float}

\usepackage{import}

\usepackage{amsmath}
\usepackage{mathtools}
\DeclarePairedDelimiter\bra{\langle}{\rvert}
\DeclarePairedDelimiter\ket{\lvert}{\rangle}
\DeclarePairedDelimiterX\braket[2]{\langle}{\rangle}{#1 \delimsize\vert #2}

\usepackage{calc}
\usepackage{color}

\usepackage{graphicx}
\usepackage{caption}
\usepackage{subfig}

\usepackage[normalem]{ulem}

\usepackage{fancyhdr}

\usepackage{url}

\usepackage[affil-it]{authblk}
\begin{document}


\title{Bifurcation in Quantum Measurement}
\author[1]{Karl-Erik Eriksson}
\author[2]{Martin Cederwall}
\author[1]{\\Kristian Lindgren\footnote{Corresponding author: kristian.lindgren@chalmers.se}}
\author[3]{Erik Sj\"oqvist}

\affil[1]{Division of Physical Resource Theory, Department of Space, 
Earth and Environment, Chalmers University of Technology, G\"oteborg, Sweden} 
\affil[2]{Division of Theoretical Physics, Department of Physics, Chalmers 
University of Technology, G\"oteborg, Sweden}
\affil[3]{Department of Physics and Astronomy, Uppsala University, Uppsala, Sweden}

\date{\today}
\maketitle

\begin{abstract}
We present a generic model of (non-destructive) quantum measurement. Being formulated within reversible quantum mechanics, the model illustrates a mechanism of a measurement process --- a transition of the measured system to an eigenstate of the measured observable. The model consists of a two-level system $\mu$ interacting with a larger system $A$, consisting of smaller subsystems. The interaction is modelled as a scattering process. Restricting the states of $A$ to product states leads to a bifurcation process: In the limit of a large system $A$, the initial states of $A$ that are efficient in leading to a final state are divided into two separated subsets. For each of these subsets, $\mu$ ends up in one of the eigenstates of the measured observable. The probabilities obtained in this branching confirm the Born rule.
\end{abstract}



\section{Introduction}

We have made a statistical study of entanglement between a two-level quantum system and a 
larger quantum system, together treated as a closed system, obeying reversible quantum 
dynamics. Applied to the problem of quantum measurement, this means that the larger system 
can be considered as part of a measurement apparatus, interacting with the two-level system.

When applying the quantum mechanics of the 1930s to this combined system, one could not see the 
possibility of transitions between the two channels corresponding to the eigenstates of the 
measured observable. Such transitions can take place due to reversibility via a return to the 
ingoing state, but the tradition that was established could not account for changes in 
the weights between the channels. 

A (non-destructive) measurement is a bifurcation process. The result is one of the eigenvalues 
of the measured observable and the system subject to measurement goes into the corresponding 
eigenstate of the measured obsevable. This was formulated early by Dirac \cite{dirac47} (p. 36):
 
\begin{quote} 
"In this way we see that a measurement always causes the system to jump into an eigenstate of 
the dynamical variable that is being measured, the eigenvalue this eigenstate belongs to being 
equal to the result of the measurement."
\end{quote}

\noindent 
In the present paper we show how an entangling interaction between the small (two-level) quantum 
system (the system subject to measurement) and the larger quantum system, through purely 
statistical mechanisms, can lead to such a bifurcation. The larger quantum system then represents 
the part of the measurement apparatus first encountered by the two-level system.

We have chosen to analyze the interaction between the two systems in the scattering theory as used in
quantum field theory, see, e.g., Ref. \cite{jauch55}. Thus, we take a holistic view of 
the entire process; as discussed in Section \ref{sec:scattering-theory} below, this leads to a non-linear dependence on the initial state.
In Section \ref{sec:model}, our model is presented in detail. It is shown how a statistical mechanism causes the scattering process to lead to one of the eigenstates and the corresponding measurement result, as described by Dirac.
In Section \ref{sec:perturbation}, it is shown how
reversibility opens for communication between the channels 
via a return to the ingoing state --- another way of seeing how a non-linear dependence on the ingoing state of the small system arises. 
The possibility for an initial state of the large system to influence 
the whole process is handled statistically under the assumption that this does not introduce any 
systematic bias. We then find that the interaction leads to the well-known 
type of bifurcation of quantum measurement governed by Born's rule. 

Another approach to the measurement process is the decoherence program with variations \cite{zeh70,zurek03,schlosshauer05}. In this approach, off-diagonal terms in the density matrix in the measured basis decay exponentially, 
while leaving the diagonal terms corresponding to the probabilities for the different 
measurement outcomes. While decoherence provides a physical mechanism for the 
appearance of probabilities in the quantum-mechanical measurement process, it fails to 
explain the appearance of definite outcomes in real experiments. 

Mathematically, the non-linearity of the bifurcation process emerging in our model is similar to the non-linearities encountered in quantum diffusion \cite{gisin84,diosi88,gisin92,percival94}.
The key point of our model in relation to quantum diffusion is that we give a quantum-mechanical explanation for how such a non-linearity may arise. Conceptually, our proposed bifurcation mechanism has some similarity with the spontaneous-collapse theory  \cite{ghirardi86,ghirardi90}, which assumes a fundamental stochastic collapse process intrinsic to each quantum-mechanical degree of freedom in nature. In order to preserve the Schr\"odinger dynamics of closed simple quantum systems, this spontaneous-collapse process must be very rare and therefore visible only in the limit of a very large number of degrees of freedom. While the spontaneous collapse theory is in effect a modified version of quantum mechanics, i.e., it would give rise to slightly different predictions in certain experiments \cite{collett03,bera15}, our model explains the bifurcation that occurs in measurement without any further modification of standard reversible quantum mechanics. In our model the non-linearity arises only as a result of normalization.

Thus, the process of measurement is not alien to quantum mechanics; its first stage is a 
quantum-mechanical closed-system evolution that has been described as a final-state interaction 
of a scattering process \cite{eriksson02,eriksson09}. Actually, reversibility is 
crucial for this process as it allows transitions between the channels via the initial state. 
With increasing size of the larger system, a bifurcation process becomes possible, leading towards 
one of the eigenstates of the measured observable. 
This is the perspective we have taken in 
the construction of the model presented here.
The main contribution with the paper is a proof of concept: a mechanism that selects one of the eigenstates can be formulated as a bifurcation process using standard quantum mechanics.

\section{Two-level quantum system interacting with a larger quantum system}
\label{two-level-system}
Consider a two-level quantum system $\mu$ in an entangling interaction with a larger quantum 
system $A$. Then $\mu$ is conveniently described in a basis given by the eigenstates of an 
operator $C$,
\begin{equation}
C \ket{\pm}_\mu = \pm \ket{\pm}_\mu \;; \; 
C = 
\begin{pmatrix} 
1 & \;0 \\
0 & \;-1 
\end{pmatrix} \;; \; 
\ket{+}_\mu = 
\begin{pmatrix} 
1  \\
0  
\end{pmatrix} \;; \,
\ket{-}_\mu = 
\begin{pmatrix} 
0  \\
1  
\end{pmatrix} \;. \label{sigma}
\end{equation}
The larger quantum system $A$ is assumed to be initially in a state
\begin{equation}
\ket{0,\alpha}_A \;. \label{state0}
\end{equation}
We think of $A$ as a part of a measurement apparatus for measuring $C$ on $\mu$. 
Then the $\mu A$-scattering is a process that leads $A$ into entanglement with $\mu$. 
The $0$ in the state of A, indicates that the interaction with $\mu$ has not yet taken place.
The unknown details of this state, represented by $\alpha$, will be described later. This process takes $\mu \cup A$ from an initial state
\begin{equation}
\ket{+}_\mu \otimes \ket{0,\alpha}_A \;\;\;\; \text{ or } \;\;\;\; \ket{-}_\mu \otimes 
\ket{0,\alpha}_A \; 
\label{initial_state}
\end{equation}
into a corresponding final state, where $\mu$ and $A$ have parted and no longer interact,
\begin{equation}
\ket{+}_\mu \otimes \ket{+,\alpha}_A \;\;\;\; \text{ or } \;\;\;\; 
\ket{-}_\mu \otimes \ket{-,\alpha}_A \; .
\label{final_state}
\end{equation}
Here the state of $A$ is labelled by the result of its interaction with $\mu$ ($+$ or $-$). 
This is clearly a process for which scattering theory is applicable.

More generally, we shall assume that initially $\mu$ has been prepared in a superposition state,
\begin{equation}
\ket{\psi}_\mu = \psi_{+} \ket{+}_\mu + \psi_{-} \ket{-}_\mu  \;, \;
( | \psi_{+} |^2 + | \psi_{-} |^2 = 1 ) \;,
\label{superposition}
\end{equation}
i.e. that for $\mu \cup A$ the combined initial state to be considered is
\begin{equation}
\ket{\psi}_\mu \otimes \ket{0, \alpha}_A = \psi_{+} \ket{+}_\mu \otimes \ket{0, \alpha}_A + \psi_{-} \ket{-}_\mu \otimes \ket{0, \alpha}_A  \; .
\label{combined_state}
\end{equation}
If one thinks of a unitary operator inducing a transition from the states in (\ref{initial_state}) 
into the states in (\ref{final_state}), then one might expect the same operator to take the 
initial state of $\mu \cup A$ into the entangled state
\begin{equation}
\psi_{+} \ket{+}_\mu \otimes \ket{+, \alpha}_A + \psi_{-} \ket{-}_\mu \otimes \ket{-, \alpha}_A  \; .
\label{entangled_state}
\end{equation}
Besides achieving the entanglement, this would lock the relative weights of the two channels at their 
initial values.

The tradition based on the final state as (\ref{entangled_state}) resulting from an entangling 
interaction between $\mu$ and the larger quantum system $A$ 
dates back to von Neumann and is well known. 
A reasoning along 
this line with fixed channel coefficients led Schr\"odinger to introduce his story of a cat in a 
superposition state of being both dead and alive. 
It also lead to Everett's relative-state formalism \cite{everett57} and its continuation 
in DeWitt's many-worlds interpretation \cite{dewitt70}. 

However, a measurement of $C$ (considering here a non-destructive measurement) is known to 
lead either to the result $+1$ {\it or} the result $-1$ and to take $\mu$ into the corresponding 
eigenstate $\ket{+}_\mu$ or $\ket{-}_\mu$, with relative frequencies $|\psi_{+}|^2$ and 
$|\psi_{-}|^2$, respectively. The disagreement of this experience with (\ref{entangled_state}) 
has led to a long discussion about quantum measurement as an extraordinary process not 
understandable within quantum mechanics itself but in need of a theory of its own. 
As far as we can see, differences in transition amplitudes have been overlooked in this discussion.

In the following section we shall use scattering theory to construct a model. This opens the 
possibility that the channels may differ in terms of the transition amplitudes taking the initial 
states in (\ref{initial_state}) into the final states in (\ref{final_state}).

\section{Scattering-theory treatment of quantum entanglement}
\label{sec:scattering-theory}
In the scattering theory related to quantum field theory, normalization in time and space forces 
one to focus on scattering amplitudes and transition rates. The transition matrix describing the 
$\mu A$-interaction takes the ingoing states in (\ref{initial_state}) into the corresponding 
outgoing states in (\ref{final_state}). 
We then have \cite{jauch55},
\begin{equation}
M \ket{+}_{\mu} \otimes \ket{0, \alpha}_A = b_{+} (\alpha) \ket{+}_\mu \otimes \ket{+,\alpha}_A  
\label{transition+}
\end{equation}
and
\begin{equation}
M \ket{-}_{\mu} \otimes \ket{0, \alpha}_A = b_{-} (\alpha) \ket{-}_\mu \otimes \ket{-,\alpha}_A , 
\label{transition-}
\end{equation}
where $M$ is the transition operator and $b_+(\alpha)$ and $b_-(\alpha)$ are scattering 
amplitudes, depending on the initial state $\ket{0, \alpha}_A$ of $A$. The states in 
Eqs.~(\ref{transition+}) and (\ref{transition-}) are the same as those in (\ref{final_state}), 
except for the weight factors $b_+(\alpha)$ and $b_-(\alpha)$. For the general ingoing state in (\ref{combined_state}) (with non-zero $\psi_{+}$ and $\psi_{-}$), we have the 
(non-normalized) outgoing state
\begin{equation}
\label{transitionM}
M \ket{\psi}_\mu \otimes \ket{0, \alpha}_A = b_+(\alpha)  \psi_{+} \ket{+}_\mu 
\otimes \ket{+, \alpha}_A + b_-(\alpha) \psi_{-} \ket{-}_\mu \otimes \ket{-, \alpha}_A  \;.
\end{equation}
Like (\ref{entangled_state}), this is an entangled state of $\mu$ and $A$, but unlike 
(\ref{entangled_state}), the transition has induced changes in the relative weights of the two 
channels, except for the unlikely case of $|b_+(\alpha)|^2=|b_-(\alpha)|^2$. However, to 
avoid bias, in our model we shall assume the means of these quantities  
over the ensemble 
of available initial states $\ket{0, \alpha}_A$ to be the same, 
\begin{equation}
\label{non-bias}
\langle\langle |b_\pm(\alpha)|^2 \rangle \rangle=g^2 \;.
\end{equation}
The role of the constant $g > 0$ will be discussed later.

\begin{figure}[htpb]
\begin{center}
\includegraphics[scale=.2]{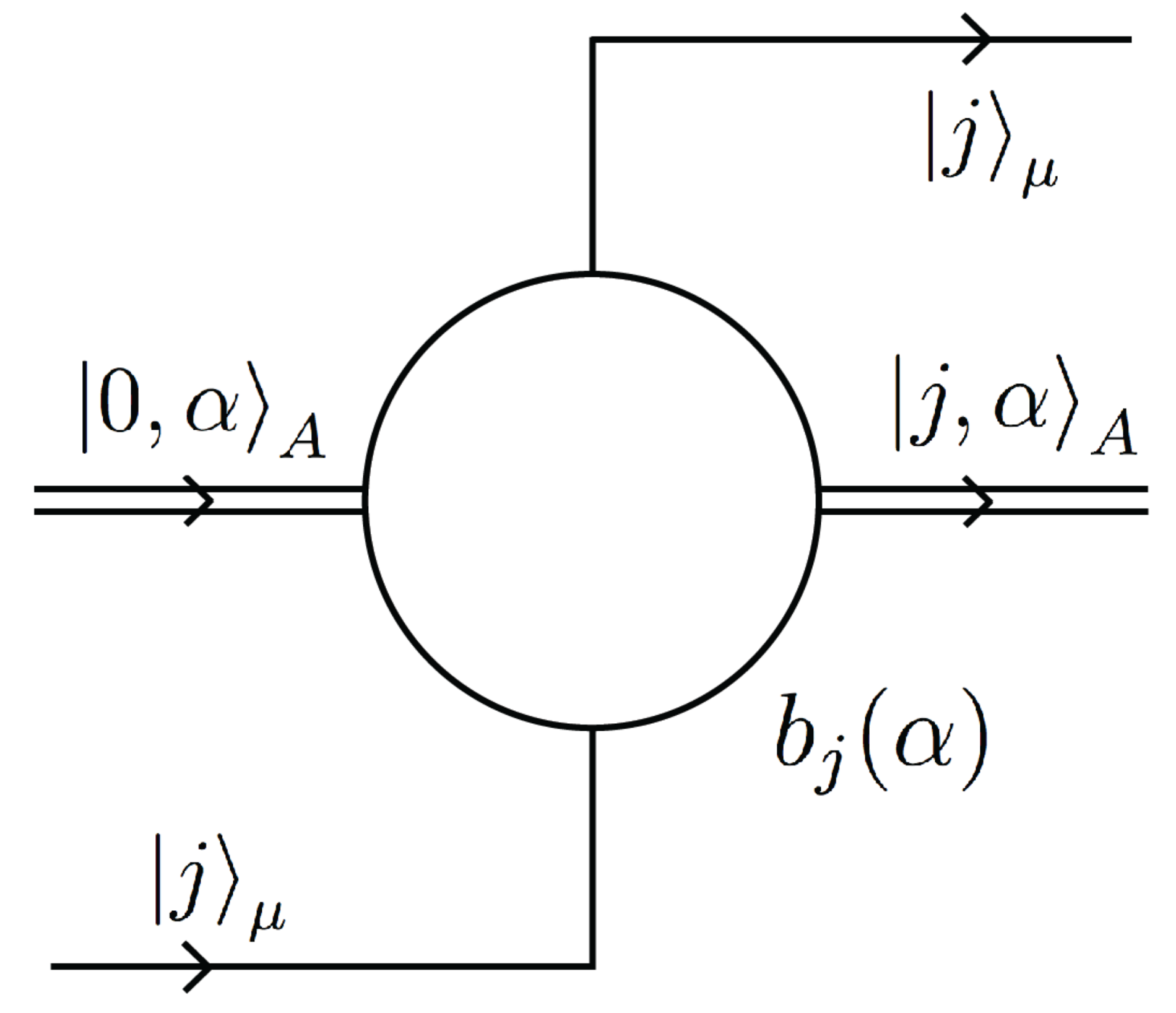}
\caption{Diagram for the transition when $\mu$ is in an eigenstate $j \in \{+,-\}$.}
\label{fig:Transition}
\end{center}
\end{figure}

\begin{figure}[htpb]
\begin{center}
\includegraphics[scale=.15]{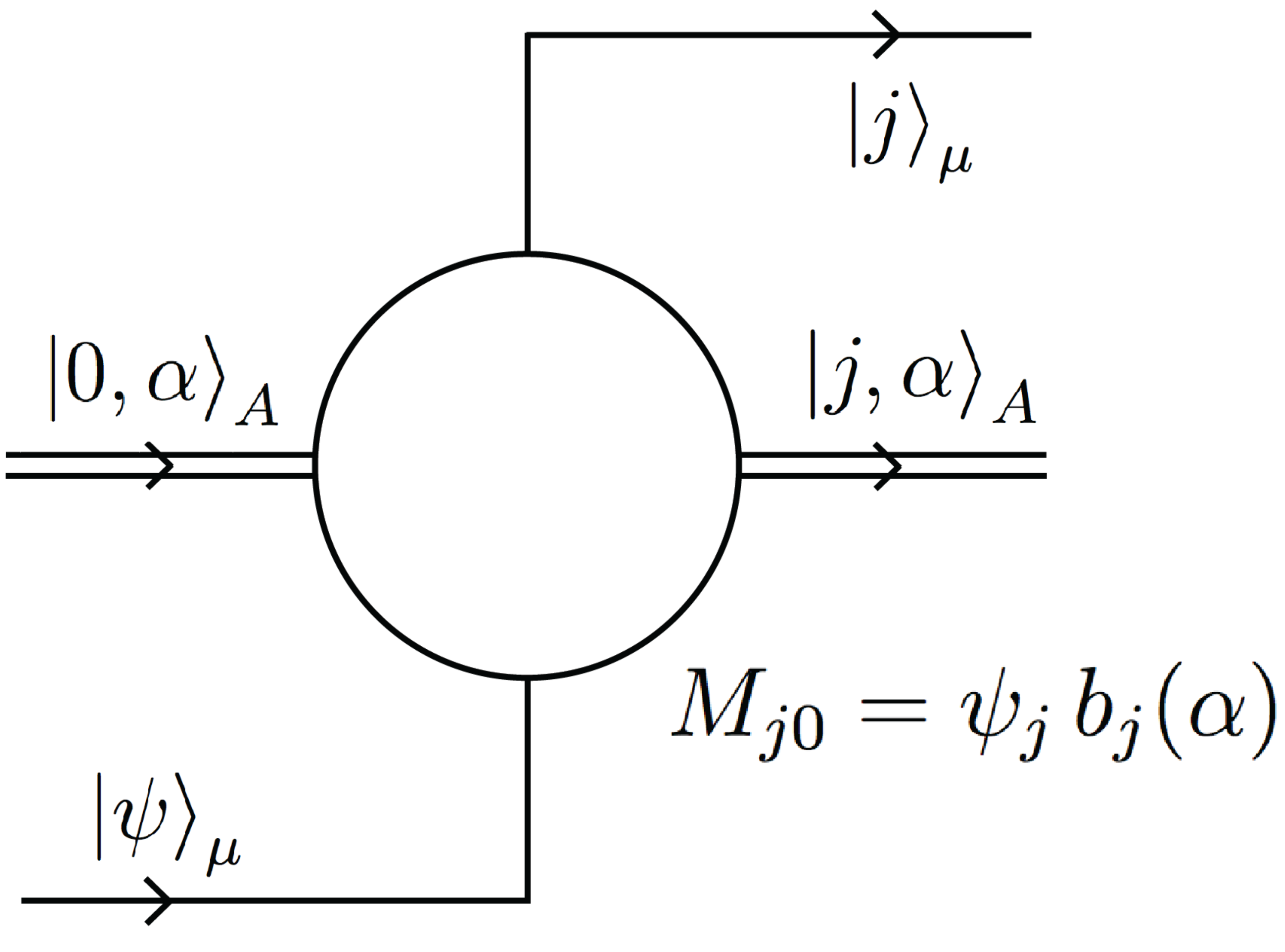}
\caption{Diagram for the transition when $\mu$ is in an initial superposition state. In $M_{j0}$, $j$ and $0$ denote the final state and the initial state, respectively.}
\label{fig:Measurement}
\end{center}
\end{figure}

The transitions described by Eqs.~(\ref{transition+}) and (\ref{transition-}) are represented by 
the diagram of Fig.~\ref{fig:Transition} with transition amplitudes $b_\pm(\alpha)$, and the 
transitions described by (\ref{transitionM}) are represented by the diagram of 
Fig.~\ref{fig:Measurement}. We limit ourselves to real and positive amplitudes 
$b_\pm(\alpha)$ in the following.

In discussions on entanglement, the dependence of transition amplitudes on the initial state 
of $A$ has usually been neglected. We shall see that differences between transition amplitudes, 
depending on the initial state of $A$, described by $\alpha$, can play a very crucial role.

Using (\ref{initial_state}) as our basis, the initial state in (\ref{combined_state}) in 
density matrix notation, is
\begin{align}
\rho_\mu^{(0)} = \ket{\psi}_{\mu \; \mu} \bra{\psi} \otimes \ket{0,\alpha}_{A \; A} \bra{0,\alpha} =
\begin{pmatrix} 
| \psi_{+} |^2  & \psi_{+} \psi_{-}^* \\[6 pt]
\psi_{-} \psi_{+}^* & | \psi_{-} |^2  
\end{pmatrix} . 
\label{s0_mtx}
\end{align}
The final state in (\ref{transitionM}) can be represented by the non-normalized projector
\begin{align}
R(\alpha) = M \big( \ket{\psi}_{\mu \; \mu} \bra{\psi} \otimes \ket{0,\alpha}_{A \; A} 
\bra{0,\alpha} \big) M^\dagger = g^2 \hat R(\alpha) \;. 
\end{align}
In the basis spanned by (\ref{final_state}), we find 
\begin{align}
\label{R}
&R(\alpha) =
\begin{pmatrix} 
| \psi_{+} |^2 b_+(\alpha)^2  & \psi_{+} \psi_{-}^* b_+(\alpha)b_-(\alpha) \\[6 pt]
\psi_{-} \psi_{+}^* b_-(\alpha)b_+(\alpha) & | \psi_{-} |^2 b_-(\alpha)^2  
\end{pmatrix}, \\[6 pt]
&R(\alpha)^2 = (\text{Tr}\; R(\alpha)) R(\alpha) \;, \nonumber 
\end{align}
see Fig.~\ref{fig:Measurement}. Apart from a numerical factor, this is the matrix of transition probabilities 
per unit time. The total transition probability per unit time is proportional to the trace
\begin{align}
\label{transitionP}
&w(\alpha) = \text{Tr}\; R(\alpha) = | \psi_+ |^2 b_+(\alpha)^2 + | \psi_- |^2 b_-(\alpha)^2 = g^2 \hat w(\alpha)\;, \text{ with} \\
&\langle\langle \hat w(\alpha) \rangle \rangle=1 \;. \nonumber
\end{align}
Thus, the normalized density matrix for the final state is
\begin{align}
\label{density_mtx}
&\rho^{(\text{f})}(\alpha) = \frac{R(\alpha)}{w(\alpha)}= \frac{\hat R(\alpha)}{\hat w(\alpha)}=  \\ 
&=\frac{1}{ | \psi_+ |^2 b_+(\alpha)^2 + | \psi_- |^2 b_-(\alpha)^2} 
\begin{pmatrix} 
| \psi_{+} |^2 b_+(\alpha)^2  & \psi_{+} \psi_{-}^* b_+(\alpha)b_-(\alpha) \\[6 pt]
\psi_{-} \psi_{+}^* b_-(\alpha)b_+(\alpha) & | \psi_{-} |^2 b_-(\alpha)^2  
\end{pmatrix} . 
\nonumber
\end{align} 
The final-state density matrix in (\ref{density_mtx}) is non-linear in the elements of the 
initial density matrix in (\ref{s0_mtx}). This is how non-linearity arises in our model (see also Section \ref{sec:perturbation}). To take the mean of (\ref{density_mtx}), 
we have to use the total transition rate in (\ref{transitionP}) as a weight. With the non-bias assumption (\ref{non-bias}), this gives us
\begin{align}
\label{rho_mean}
\langle \rho^{(\text{f})} \rangle  &=  \langle\langle \hat w(\alpha) \rho^{(\text{f})}(\alpha) \rangle\rangle = 
\langle\langle \hat R(\alpha) \rangle\rangle = \\[6 pt]
&= \begin{pmatrix} 
| \psi_{+} |^2  & \frac{\psi_{+} \psi_{-}^*}{g^2} \langle\langle b_+(\alpha)b_-(\alpha)  \rangle\rangle \\[6 pt]
\frac{\psi_{-} \psi_{+}^*}{g^2} \langle\langle b_-(\alpha)b_+(\alpha)  \rangle\rangle & | \psi_{-} |^2   
\end{pmatrix} . \nonumber
\end{align}
Thus linearity is restored. 
Moreover, we shall see that in our model, the non-diagonal matrix elements are small and we get the expected result for the statistical mean.

To reproduce the situation described by Dirac as quoted above, (\ref{density_mtx}) should lead either to 
$\rho^{(\text{f})}=
\begin{psmallmatrix}
1 & 0 \\ 0 & 0
\end{psmallmatrix}
$
or to
$\rho^{(\text{f})}=
\begin{psmallmatrix}
0 & 0 \\ 0 & 1
\end{psmallmatrix}
$.
This is possible if either $b_+>>b_-$ or $b_->>b_+$. We shall see that in our model, the final states will always result in one of these cases.

The non-linear dependence on $\rho^{(0)}$, defined in (\ref{s0_mtx}), of the final-state density 
matrix $\rho^{(\text{f})}$ in (\ref{density_mtx}) can be looked at in different ways. Here it has entered as the 
transition-rate matrix, normalized through division by the total transition rate. Equivalently, 
the density matrix can be seen as the matrix of conditional probabilities for final states, provided 
a transition to a final state has taken place. Another way to derive the non-linearities of (\ref{density_mtx}) is to include renormalization effects due to repeated returns to the 
initial state, thus taking reversibility into account. This will be shown in Section \ref{sec:perturbation}. 

\section{A model for step-wise increase of system size}
\label{sec:model}
To analyze scattering between the two systems $\mu$ and $A$,
we carry out a repeated mapping in small steps numbered $n=1, 2, ..., N$, from no scattering, increasing the part of $A$ involved in the scattering, 
along  $A^{(1)} \subset A^{(2)} \subset ... \subset A^{(N)} = A$, 
to the scattering between $\mu$ and the whole of $A$. Here $A$ is considered to consist of $N$ separate, and initially independent, subsystems $A_1, A_2, ..., A_N$, and 
\begin{equation}
A^{(n)} = \bigcup_{m=1}^n A_m \;.
\end{equation}
The $n$th step of the mapping extends the system considered from $\mu \cup A^{(n-1)}$ to $\mu \cup A^{(n)}$.

We let the symbol $\alpha$ already used to characterize the initial state of $A$, denote $N$ independent symbols $\alpha_n$ ($n=1, 2, ..., N$), characterizing the initial states of the corresponding subsystems,
\begin{equation}
\alpha =(\alpha_1, \alpha_2, ..., \alpha_N) \;.
\end{equation}
The initial state of $A^{(n)}$ is then assumed to be a product state,
\begin{align}
\ket{0, \alpha^{(n)}}_{A^{(n)}} = \; &\ket{0; \alpha_1}_{A_1} \otimes 
\ket{0; \alpha_2}_{A_2}  \otimes ... \otimes \ket{0; \alpha_n}_{A_n} \; , \\
& \text{with } \; \alpha^{(n)}=(\alpha_1, \alpha_2, ..., \alpha_n) \;. \nonumber
\label{n-state}
\end{align}
In the $n$th step, we go from considering the initial state of $\mu \cup A^{(n-1)}$,
\begin{equation}
\ket{\psi}_{\mu} \otimes \ket{0; \alpha^{(n-1)}}_{A^{(n-1)}}
\end{equation}
to the initial state of $\mu \cup A^{(n)}$,
\begin{equation}
\ket{\psi}_{\mu} \otimes \ket{0; \alpha^{(n)}}_{A^{(n)}} =
\ket{\psi}_{\mu} \otimes \ket{0; \alpha^{(n-1)}}_{A^{(n-1)}} \otimes
\ket{0; \alpha_n}_{A_n} \;.
\end{equation}
The related final states after scattering within $\mu \cup A^{(n)}$ are the states
\begin{equation}
\ket{j}_{\mu} \otimes \ket{j; \alpha^{(n)}}_{A^{(n)}} =
\ket{\psi}_{\mu} \otimes \ket{j; \alpha^{(n-1)}}_{A^{(n-1)}} \otimes
\ket{j; \alpha_n}_{A_n} \;.
\end{equation}
In the $n$th step, we assume the transition amplitudes to acquire new factors depending on the $n$th subsystem $A_n$, as follows,
\begin{align}
\label{model-statistics}
&b_\pm^{(n)} = b_\pm^{(n-1)} g_n (1 \pm \tfrac{1}{2} \eta_n - \tfrac{1}{8} \kappa_n^{\;2} )  ; \;\; \eta_n=\eta_n(\alpha_n)  ; \;\; \eta_n = \eta_n^* \;; \\
&\langle \langle \eta_n(\alpha_n) \rangle \rangle = 0 \;, \;\; 
\langle \langle \eta_n(\alpha) \eta_{n'}(\alpha) \rangle \rangle = \delta_{n n'} \kappa_n^{\;2}  : \;\; \kappa_n^{\;2} <<1 \;. \nonumber
\end{align}
For each step, we thus keep terms up to second order in $\eta_m$ and use the convention to replace second-order terms by their mean values. The relations for the bilinear forms of the amplitudes are
\begin{align}
&b_\pm^{(n)2} = b_\pm^{(n-1)2} g_n^{\;2} (1 \pm \eta_n ) \;, \\
&b_+^{(n)} b_-^{(n)} = b_+^{(n-1)} b_-^{(n-1)} g_n^{\;2} (1-\tfrac{1}{2} \kappa_n^{\;2}) \;.
\nonumber
\end{align}
For the amplitudes involving the whole system $A$, we get
\begin{align}
\label{b-Y}
& b_\pm(\alpha)^2 = \prod_{n=1}^N g_n^{\;2} (1\pm \eta_n(\alpha_n)) = 
g^2 e^{\Xi(\pm Y-\tfrac{1}{2})} \;, \nonumber \\
& b_+(\alpha) b_-(\alpha) = g^2 e^{-\tfrac{1}{2}\Xi} \;; \\
& g = \prod_{n=1}^N g_n \;,\;\; Y=Y(\alpha) = \frac{1}{\Xi} \sum_{n=1}^N \eta_n(\alpha_n) \;, \;\; \Xi = \sum_{n=1}^N \kappa_n^2 \;. \nonumber
\end{align}
Here $\Xi$ is the total step variance and $Y$ is an accumulated variable depending on details of $\alpha$, describing the initial state $\ket{ 0, \alpha }_A$ of $A$. $Y$ is clearly a variable causing enhancement of one channel and suppression of the other. For the relevant mean values we get
\begin{align}
\label{Y-prop}
\langle \langle Y \rangle \rangle = 0 \;,\;\; 
\big\langle \big\langle Y^2 \big\rangle \big\rangle = \frac{1}{\Xi}  \;,\;\; 
\Big\langle \Big\langle e^{\Xi(\pm Y-\tfrac{1}{2})} \Big\rangle \Big\rangle = 1 \;. 
\end{align}
We shall use the total variance $\Xi$ rather than $N$ as a measure of the extension of the system $A=A^{(N)}$.

The transition-rate matrix, or the final-state non-normalized density matrix, (\ref{R}) is
\begin{align}
\label{R(Y)}
R(Y)=g^2 \hat R(Y)\;, \;\; \hat R(Y) = e^{\,-\tfrac{1}{2}\Xi} \,
\begin{pmatrix} 
| \psi_{+} |^2 e^{\Xi Y} & \psi_{+} \psi_{-}^* \\[6 pt]
\psi_{-} \psi_{+}^* & | \psi_{-} |^2 e^{-\Xi Y} 
\end{pmatrix} \;,
\end{align}
and its trace is
\begin{align}
\label{w(Y)}
w(Y) = g^2 \hat w(Y) \;,\;\; \hat w(Y) = e^{-\tfrac{1}{2} \Xi} \big( | \psi_{+} |^2 e^{\Xi Y} + 
| \psi_{-} |^2 e^{-\Xi Y} \big) \;.  
\end{align}
Here, $w(Y)$ is the squared norm of the state in (\ref{transitionM}) and a measure of the total 
transition rate. Together with (\ref{Y-prop}), equation (\ref{w(Y)}) yields the ensemble mean of $\hat w(Y)$  
\begin{align}
\langle\langle \hat w(Y) \rangle\rangle = 1 \;. 
\end{align}
Using (\ref{R(Y)}) and (\ref{w(Y)}), we find for the normalized final-state density matrix 
in (\ref{density_mtx}),
\begin{align}
\label{rho(Y)}
\rho^{(\text{f})}(Y) = \frac{\hat R(Y)}{\hat w(Y)} = \frac{1}{| \psi_{+} |^2 e^{\Xi Y} + | \psi_{-} |^2 e^{-\Xi Y}}
\begin{pmatrix} 
| \psi_{+} |^2 e^{\Xi Y} & \psi_{+} \psi_{-}^* \\[6 pt]
\psi_{-} \psi_{+}^* & | \psi_{-} |^2 e^{-\Xi Y} 
\end{pmatrix} . 
\end{align}
It still describes a pure state (no decoherence!),
\begin{align}
\rho^{(\text{f})}(Y)^2 = \rho^{(\text{f})}(Y) \;.
\end{align}
To find the mean of (\ref{rho(Y)}), we use $\hat w(Y)$ as a weight. This gives a 
more explicit version of (\ref{rho_mean}),
\begin{align}
\label{rho_w}
\langle \rho^{(\text{f})} \rangle = 
\langle\langle \hat w(Y) \rho^{(\text{f})}(Y) \rangle\rangle =
\langle\langle \hat R(Y) \rangle\rangle = 
\begin{pmatrix} 
| \psi_{+} |^2  & e^{- \tfrac{1}{2}\Xi Y}\psi_{+} \psi_{-}^* \\[6 pt]
e^{-\tfrac{1}{2}\Xi Y}\psi_{-} \psi_{+}^* & | \psi_{-} |^2  
\end{pmatrix}
\end{align}
In the ensemble of initial states, the distribution over $Y$ obtained from (\ref{b-Y}) is
\begin{align}
q(Y) = \Big\langle \Big\langle \delta\Big( Y - \sum_{n=1}^N \eta_n(\alpha_n) \Big) \Big\rangle \Big\rangle \;.
\end{align}
For sufficiently large $\Xi$ this is well approximated by
\begin{align}
\label{phase-space-factor}
q(Y) = \sqrt{ \frac{\Xi}{2\pi} } \, e^{-\tfrac{1}{2}\Xi Y^2} \;.
\end{align}
Because of the dependence of the transition amplitudes on the initial state of $A$, $\ket{0, \alpha}_A$, through 
$Y$ as given in (\ref{b-Y}), the initial states vary strongly in their efficiency to lead to a 
transition. 
Beside the phase space factor (\ref{phase-space-factor}), the 
distribution $Q(Y)$ of final states includes the normalized transition rate $\hat w(Y)$ as a factor 
\begin{align}
\label{Q(Y)}
&Q(Y)= q(Y)  \hat w(Y)= | \psi_+ |^2 Q_+(Y) + | \psi_- |^2 Q_-(Y) \; ;  \\
&Q_\pm(Y) = \sqrt{ \frac{\Xi}{2\pi} } \, e^{-\tfrac{1}{2}\Xi (Y\mp 1)^2} . 
\nonumber
\end{align}
This implies a strong selection among the available initial states.
At the peaks at $Y=\pm 1$, the density matrix in (\ref{rho(Y)}) of the final state is
\begin{align}
\label{rho(1)}
\rho^{(\text{f})}(1) &= \Big( 1 + e^{-2\Xi} \Big| \frac{\psi_-}{\psi_+} \Big|^2 \, \Big)^{-1}
\begin{pmatrix} 
1  & e^{-\Xi}\frac{\psi_-^*}{\psi_+^*} \\[8 pt]
e^{-\Xi}\frac{\psi_-}{\psi_+}  & e^{-2\Xi} \Big| \frac{\psi_-}{\psi_+} \Big| ^2  
\end{pmatrix} , \\[8 pt]
\label{rho(-1)}
\rho^{(\text{f})}(-1) &= \Big( 1 + e^{-2\Xi} \Big| \frac{\psi_+}{\psi_-} \Big|^2 \, \Big)^{-1}
\begin{pmatrix} 
e^{-2\Xi} \Big| \frac{\psi_+}{\psi_-} \Big| ^2  & e^{-\Xi}\frac{\psi_+}{\psi_-} \\[8 pt]
e^{-\Xi}\frac{\psi_+^*}{\psi_-^*}   & 1  
\end{pmatrix} \; .
\end{align}
In Section \ref{sec:large-system}, we shall discuss the limit of large $\Xi$ more in detail.

\begin{figure}[htpb]
\begin{center}
\includegraphics[scale=.65]{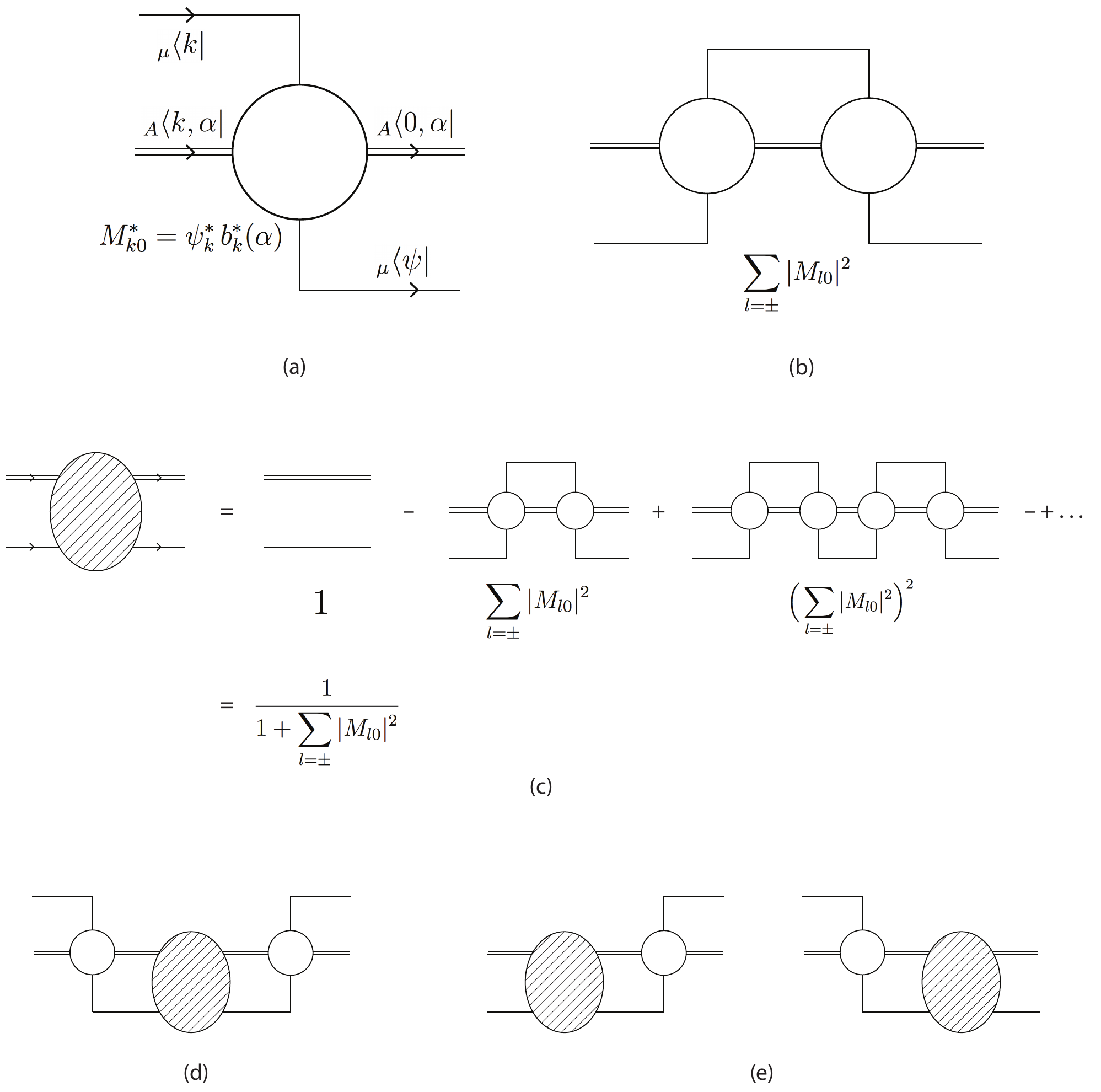}
\caption{Feynman diagrams. (a) Diagram for the Hermitean conjugate of the transition ($M^*_{k0}$) to $k=\pm$.
(b) Second-order diagram for no change ($0 \rightarrow 0$).
(c) Diagrams to all orders for no change; this can also be interpreted as renormalization of the initial state.
(d) Sum over diagrams to all orders for the $jk$-component ($j,k=\pm$) of the final-state density matrix.
(e) Diagrams for the $j0$- and the $0k$-components of the final-state density matrix.}
\label{fig:scattering}
\end{center}
\end{figure}

\section{Perturbation expansion of the scattering process}
\label{sec:perturbation}
The inverse process of the scattering in Fig.~\ref{fig:Measurement} is represented by the diagram 
of Fig.~\ref{fig:scattering}a. The two taken together, as shown in Fig.~\ref{fig:scattering}b, represent 
a loss from the initial state due to scattering. Fig.~\ref{fig:scattering}b can be repeated any number 
of times. The sum over diagrams of the initial state going into itself, Fig.~\ref{fig:scattering}c, forms 
a geometrical series, representing the probability for the initial state to remain unchanged, i.e.,
\begin{align}
\label{00-transition}
\frac{1}{1+|\psi_+|^2 \, b_+^2+|\psi_-|^2 \, b_-^2} = \frac{1}{1+g^2 \, \hat w(Y)} \;.
\end{align}
Thus, the scattering probability is
\begin{align}
\label{p-scattering}
\frac{g^2 \, \hat w(Y)}{1+g^2 \, \hat w(Y)} \;.
\end{align}
Consider a $(3\times3)$-matrix representation of the final state, that also includes the initial state, we can label it $0$, and allows $0 \rightarrow 0$ transitions. Then the probability for the initial state to remain unchanged (\ref{00-transition}) is the $(0,0)$-element, and the scattering probability (\ref{p-scattering}) is the trace of the $(2\times2)$-submatrix of scattering states, represented by Fig. \ref{fig:scattering}d. Diagrams corresponding to the off-diagonal elements involving 0, are shown in Fig. \ref{fig:scattering}e.

The final-state $(3\times3)$ density matrix expressed in the orthonormal basis 
$\{ \ket{\psi}_{\mu} \otimes \ket{0,\alpha}_A, \ket{+}_{\mu} \otimes \ket{+,\alpha}_A, 
\ket{-}_{\mu} \otimes \ket{-,\alpha}_A \}$ is
\begin{align}
\label{3x3-matrix}
\bar{\rho}^{(\text{f})} &= \frac{1}{g^{-2} e^{\tfrac{1}{2}\Xi} + | \psi_{+} |^2 e^{\Xi Y} + | \psi_{-} |^2 e^{-\Xi Y}} \times \\[12 pt] \nonumber
&\times
\begin{pmatrix} 
g^{-2} e^{\tfrac{1}{2}\Xi}    &    g^{-1} e^{\tfrac{1}{2}\Xi (Y+\tfrac{1}{2})}\psi_+^*    &   g^{-1} e^{\tfrac{1}{2}\Xi (-Y+\tfrac{1}{2})}\psi_-^*      \\[6 pt]
g^{-1} e^{\tfrac{1}{2}\Xi (Y+\tfrac{1}{2})}\psi_+    &    | \psi_{+} |^2 e^{\Xi Y}        &    \psi_{+} \psi_{-}^* \\[6 pt]
g^{-1} e^{\tfrac{1}{2}\Xi (-Y+\tfrac{1}{2})}\psi_-    &    \psi_{-} \psi_{+}^*               &    | \psi_{-} |^2 e^{-\Xi Y} 
\end{pmatrix} \;.
\end{align}
This is the result of a unitary S-matrix acting on the initial density operator in (\ref{combined_state}) 
as described by the diagrams of Fig. 3c, 3d and 3e,
\begin{align}
\label{unitary}
\bar{\rho}^{(\text{f})} = S
\begin{pmatrix} 
1  &    \;\;0    &  \;\;0     \\[3 pt]
0  &    \;\;0    &  \;\;0     \\[3 pt]
0  &    \;\;0    &  \;\;0  
\end{pmatrix}
S^\dagger \;.
\end{align}

This method of diagrammatic representation of the bilinear form of scattering amplitudes and 
their complex conjugates was used long ago by Nakanishi \cite{nakanishi58} to describe soft-photon 
emission in quantum electrodynamics.

In the strong-coupling limit, $g \rightarrow \infty$, in (\ref{3x3-matrix}), the $0 \rightarrow 0$ 
transitions disappear and the state can be fully represented by the $(2\times2)$-submatrix (\ref{rho(Y)}).

\section{The limit of a large system $A$}
\label{sec:large-system}
The combined system $\mu \cup A = \mu \cup A^{(N)}$ is assumed to develop according to 
reversible quantum mechanics. Then $A$ should not be too large.
Still, it should be possible to have the 
total step variance $\Xi=\sum_{n=1}^N \kappa_n^{\;2}$
sufficiently large. 
Bell has given a principle concerning the limit of the purely quantum-mechanical treatment
\cite{heisenberg30,bell04}, i.e., 
the size of $A$:\\[-8pt]

\begin{quote}
"... put sufficiently much into the quantum system that the inclusion of more would not 
significantly alter practical predictions."
\end{quote}

\noindent
In our case, looking at the final-state distribution $Q(Y)$ in (\ref{Q(Y)}) and its peaks at $Y=\pm1$, 
we see that the crucial quantity is $e^{-\Xi}=\prod_{n=1}^N (1-\kappa_n^{\;2})$.
Thus if the total variance $\Xi$ 
is large enough for $e^{-\Xi}$ to be negligibly small, then we have followed Bell's principle 
and included sufficiently much in $A=A^{(N)}$.

In the limit of large $\Xi$, the distribution of final states (\ref{Q(Y)}) takes the form 
\begin{eqnarray}
\label{final-states-distr}
Q(Y) = | \psi_+ |^2 \delta (Y - 1) + | \psi_- |^2 \delta (Y + 1) 
\end{eqnarray}
yielding the final density matrices (\ref{rho(1)}) and (\ref{rho(-1)}),
\begin{align}
\label{Q_final1}
&\rho^{(\text{f})}(1) =
\begin{pmatrix} 
1  &  \;0 \\
0  &  \;0
\end{pmatrix} \; , \;
\rho^{(\text{f})}(-1) =
\begin{pmatrix} 
0  &  \;0 \\
0  &  \;1
\end{pmatrix} 
\end{align}
and the overall mean 
\begin{align}
\label{Q_final2}
&\langle\langle \hat w(Y) \rho^{(\text{f})}(Y) \rangle\rangle = 
\begin{pmatrix} 
| \psi_+ | ^2  & 0 \\[6 pt]
0  & | \psi_- | ^2 
\end{pmatrix} .
\end{align}
Note that the final states distribution (\ref{final-states-distr}) reflects a strong selection of states in the initial ensemble --- states that have a high transition rate. In the following discussion we need only consider such initial states. Thus, in the limit of large $\Xi$, in such a state,
$\mu$ goes either into $\ket{+}_{\mu \; \mu} \bra{+}$ or into
$\ket{-}_{\mu \; \mu} \bra{-}$, i.e., one term in (\ref{transitionM}) becomes negligible, 
entanglement ceases and we reach again a {\em product state} for $\mu \cup A$. 

\begin{figure}[htpb]
\begin{center}
\includegraphics[scale=.3]{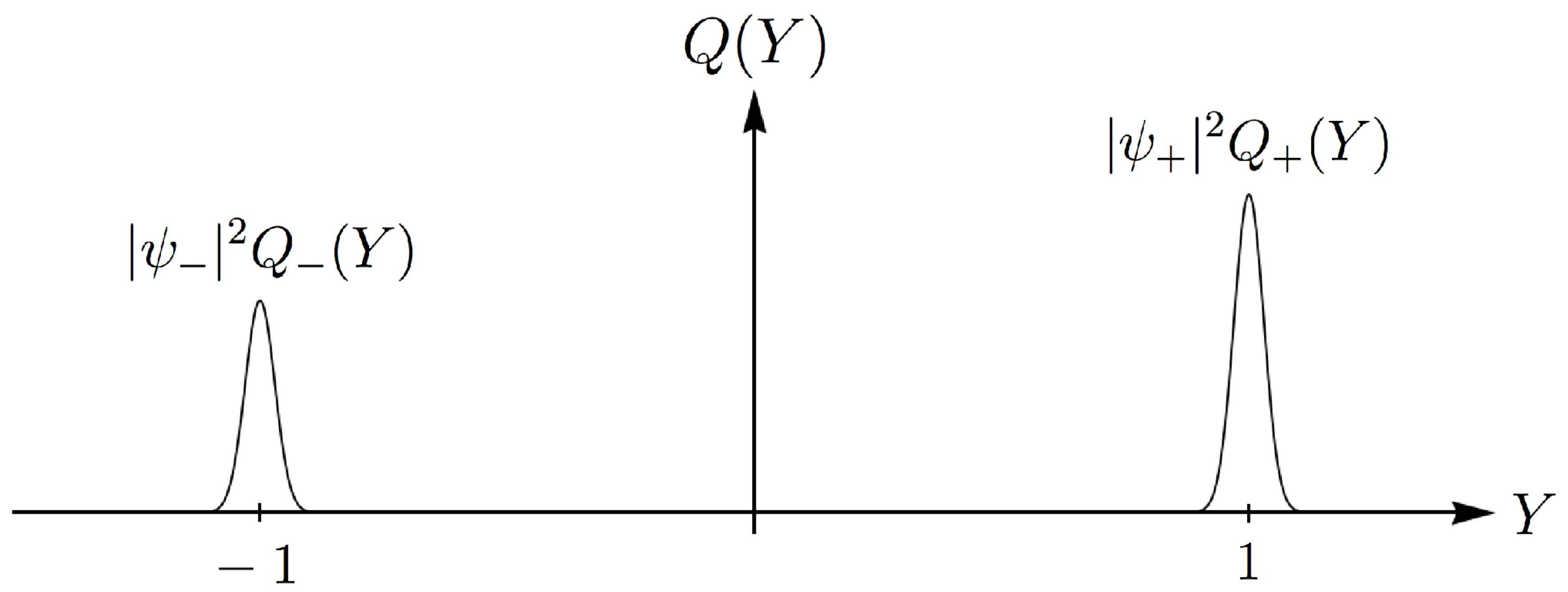}
\caption{The distribution $Q(Y)$ of final states over the aggregate quantity $Y$.}
\label{fig:Bifurcation}
\end{center}
\end{figure}

\begin{figure}[htpb]
\begin{center}
\includegraphics[scale=.3]{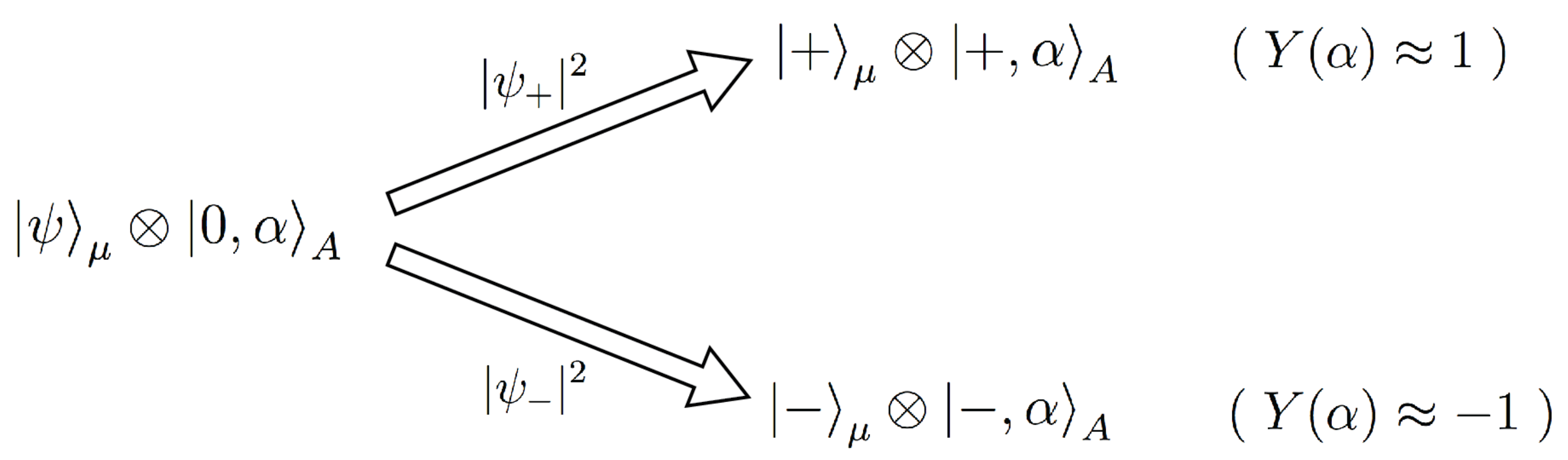}
\caption{Schematic diagram of the bifurcation resulting from $\mu A$-interaction. 
(The ingoing state $\ket{0,\alpha}_A$ is among those states that lead to a fast transition.)}
\label{fig:Bifurcation-scheme}
\end{center}
\end{figure}

Thus the initial 
state (\ref{combined_state}), which for $\mu$ is the superposition (\ref{superposition}), a 
'{\em both-and} state', ends up with $\mu$ {\em either} in $\ket{+}_{\mu \; \mu} \bra{+}$ 
{\em or} in $\ket{-}_{\mu \; \mu} \bra{-}$, depending on whether the initial state of $A$ 
in (\ref{state0}) happens to be in the subensemble with $Y(\alpha)$ at the $+1$ peak 
or in the subensemble with $Y(\alpha)$ at the $-1$ peak of Fig. \ref{fig:Bifurcation}, narrowed 
to delta functions. A scheme for this bifurcation is shown in Fig. \ref{fig:Bifurcation-scheme}.

Moreover, the relative weights for these peaks, or subensembles, are $| \psi_{+} |^2$ and 
$| \psi_{-} |^2$, respectively, in agreement with the {\em Born rule}.

In (\ref{rho(Y)}), we can follow the transition from the initial state in (\ref{s0_mtx}) 
from $\Xi=0$ (no interaction yet between $\mu$ and $A$) along increasing $\Xi$ to the 
asymptotic final states $\rho^{(\text{f})}(1)$ and $\rho^{(\text{f})}(-1)$ in (\ref{Q_final1}) for large $\Xi$.

Similarly, (\ref{Q(Y)}) interpolates between a situation with small $\Xi$, and a situation with 
large $\Xi$, which means from a broad unimodal distribution over $Y$ with a maximum at 
$Y=0$ for small $\Xi$ to a bimodal distribution with sharp separated maxima at $Y=\pm1$ for large $\Xi$. 

\section{Conclusions}
\label{sec:conclusions}
\noindent
We have examined a small two-level system in an entangling interaction with a larger quantum system. We have demonstrated how a bifurcation takes place through a purely statistical mechanism, as a consequence of the large number of degrees of freedom of the larger system. The many factors of enhancement and suppression lead to a strong selection among the states in the initial ensemble. As mentioned in Section \ref{sec:scattering-theory}, the mathematical treatment here is close both to quantum diffusion and to spontaneous-collapse theory. The key difference is that we describe the whole process within standard quantum mechanics.

The main contribution with the presented model is that of a proof of concept: It is indeed possible to formulate within standard quantum mechanics a model mechanism that describes a measurement process. 
The smaller system is considered to be subject to measurement and the larger system is the part of the measurement apparatus that it first encounters. 
The larger system is assumed to be initially in a state which is the product of the states for a large number of independent subsystems. We have then shown
how a definite outcome of a quantum measurement occurs within standard reversible quantum mechanics as the result of a specific
configuration of the apparatus. 
This configuration is formed randomly, as described by the statistics of the model (\ref{model-statistics}), and selected by transition rates (\ref{Q(Y)}).
This result can be generalized by increasing the number of available states of the smaller system, leading to several branches, each corresponding to a possible measurement outcome.

We emphasize that reversible quantum mechanics is not only compatible with the bifurcation in measurement but it is essential for the bifurcation to take place. In Section \ref{sec:perturbation}, it was shown how reversibility, as it manifests itself in a perturbation expansion, leads to what looks like a competition between the channels where one channel always wins, i.e. one definite measurement result always obtains.

\section{Acknowledgements}
Financial support from The Royal Society of Arts and Sciences in Gothenburg was important for the establishment of our collaboration. K.-E.E. thanks The University of Cape Coast and Karlstad University for hospitality during an early phase of this project. The present hospitality at Chalmers University of Technology is also gratefully acknowledged. K.-E.E. is also grateful for discussions with several colleagues on various aspects of the measurement problem: Marcus Berg, Patrick Dorey, Magdalena Eriksson, Bengt Gustafsson, Gunnar Ingelman, Tomas K\aa berger, Ingvar Lindgren, Bengt Nord\'en, Kazimierz Rz\c{a}{\.z}ewski, Per Salomonsson, Bo-Sture Skagerstam and Bo Sundborg. E.S. acknowledges support from the Swedish Research Council (VR) through Grant No. D0413201.



\end{document}